\documentclass[11pt]{article}
\usepackage{hyperref}
\usepackage{latexsym}
\usepackage{amssymb}
\usepackage{epsfig}
\usepackage{rotating}
\usepackage{amsfonts}
\usepackage{amsmath}
\usepackage{graphicx}
\newcommand\pngright[4]{
\Zw=#2 \divide \Zw by 5
\Zh=#3 \divide \Zh by 60  \advance\Zh by 1
\setcounter{wrapwidth}{\Zw}
\begin{wrapfigure}[\Zh]{r}{\value{wrapwidth}pt}
\begin{center}
\vspace{#4pt}
\includegraphics*[width=\Zw pt]{images/#1}
\end{center}
\end{wrapfigure}}

\newlength{\Taille}

\input xy
\xyoption{all}

\newcommand{\flechebas}[1]{
  \settoheight{\unitlength}{\mbox{$#1$}}
  \settowidth{\Taille}{\mbox{~${\scriptstyle #1}$}}
  \addtolength{\unitlength}{4ex}
  \begin{picture}(0,1)
    \put(0,1){\vector(0,-1){1}}
    \put(0,0.5){\makebox(0,0){${\scriptstyle #1}$ \hspace{\the\Taille}}}
  \end{picture}}
\newcommand{\flechehaut}[1]{
  \settoheight{\unitlength}{\mbox{$#1$}}
  \settowidth{\Taille}{\mbox{~${\scriptstyle #1}$}}
  \addtolength{\unitlength}{4ex}
  \begin{picture}(0,1)
    \put(0,0){\vector(0,1){1}}
    \put(0,0.5){\makebox(0,0){\hspace{\the\Taille}${\scriptstyle #1}$ }}
  \end{picture}}
\newcommand{\flechedroite}[1]{
  \settowidth{\unitlength}{\mbox{$#1$}}
  \settoheight{\Taille}{\mbox{${\scriptstyle #1}$}}
  \addtolength{\Taille}{1ex}
  \addtolength{\unitlength}{4ex}
  \raisebox{0.5ex}{
  \begin{picture}(1,0)
    \put(0,0){\vector(1,0){1}}
    \put(0.5,0){\makebox(0,0){${\scriptstyle #1}$ \vspace{\the\Taille}}}
  \end{picture}}}
\newcommand{\flechegauche}[1]{
  \settowidth{\unitlength}{\mbox{$#1$}}
  \settoheight{\Taille}{\mbox{${\scriptstyle #1}$}}
  \addtolength{\Taille}{1ex}
  \addtolength{\unitlength}{4ex}
  \raisebox{0.5ex}{
  \begin{picture}(1,0)
    \put(1,0){\vector(-1,0){1}}
    \put(0.5,0){\makebox(0,0){${\scriptstyle #1}$ \vspace{\the\Taille}}}
  \end{picture}}}

\newtheorem{proposition}{Proposition}[section]

\newtheorem{remark}{Remark}[section]

\begin{document}
\begin{titlepage}
\begin{flushright}
{\it \bf ICMPA-MPA/007/2012}
\end{flushright}
\begin{center}
{\LARGE \bf {On the harmonic oscillator properties in a twisted Moyal plane}}\\
\vspace{1cm}
 Dine Ousmane Samary$^{1,*}$, Mahouton Norbert Hounkonnou$^{1,\dag}$ and Ezinvi Baloitcha$^{1,\nmid}$

 $^{1}${\em University of Abomey-Calavi,\\
International Chair in Mathematical Physics
and Applications}\\
{\em (ICMPA--UNESCO Chair), 072 B.P. 50  Cotonou, Republic of Benin}\\

E-mails:  $^{*}$dine.ousmanesamary@cipma.uac.bj,\\
 $^{\dag}$norbert.hounkonnou@cipma.uac.bj,\\
$^{\nmid}$ezinvi.baloitcha@cipma.uac.bj.\\

\begin{abstract}
This work prolongs, using  an operator method, the investigations started in our recent paper 
[{\rm \it  J. Math. Phys. {\bf 51}
102108}]  on the spectrum and  states of the harmonic
oscillator on twisted Moyal plane, where rather
 a Moyal-star-algebraic approach was used.  The physical spectrum and  states of
the harmonic oscillator  on twisted Moyal space,  obtained here
 by solving the corresponding differential equation, are similar to  those  of the ordinary Moyal space, with different parameters. This fortunately contrasts with the previous study which produced unexpected results, i.e.  infinitely
degenerate states with energies depending on the coordinate functions.
\end{abstract}


\end{center}
\end{titlepage}

\section{Introduction}
The harmonic oscillator is undoubtedly one of the most important model systems in classical and quantum physics,
what justifies its intense study in the literature. Referring the reader to our previous work \cite{hd2}
 for the motivations of this investigation, we briefly recall here some salient features needed to understand 
our development in the sequel.
The noncommutative (NC) spacetime is described by the commutation relation between coordinate operators as
\begin{eqnarray}
[\hat{x}^\mu,\hat{x}^\nu]=i\widetilde{\Theta}^{\mu\nu}(x).
\end{eqnarray}
In this equation $\widetilde{\Theta}^{\mu\nu}$ is a deformation parameter tensor  which
should  vanish at large distances where one experiences
the commutative world and may be determined by experiments in the high energy case. In the simpler case when
$\widetilde{\Theta}^{\mu\nu}(x)\equiv \Theta^{\mu\nu}$, where $\Theta^{\mu\nu}$ is a constant skew-symmetric tensor, the
NC spacetime remains flat. Noncommutative field theory defined on this $\Theta-$deformed spacetime is extensively studied 
in the literature  \cite{aschieri}-\cite{wess2}. When the parameter
$\widetilde{\Theta}^{\mu\nu}(x)$ depends on the space variables, it can  engender a dynamical twisted space and
the geometry associated with this deformation becomes curve.
Consider, (see \cite{Goursac1} and references therein), $E=\{\hat{x}^\mu, \mu\in [[1,2,\cdots D]]\}$ and
$\mathbb{C}[[\hat{x}^{1}, \hat{x}^{2},\cdots \hat{x}^D]],$ the free algebra generated by $E$.
Let $\mathcal{I}$ be the ideal of $\mathbb{C}[[x^{1}, x^{2}, \cdots x^D]],$
engendered by the elements $\hat{x}^\mu \hat{x}^\nu-\hat{x}^\nu \hat{x}^\mu
-i\tilde{\Theta}^{\mu\nu}$. The twisted Moyal Algebra
$\hat{\mathcal{A}}_{\tilde\Theta}$ is the quotient
\begin{eqnarray}
\mathbb{C}[[\hat{x}^{1},
\hat{x}^{2},\cdots \hat{x}^D]]/\mathcal{I}.
\end{eqnarray}
 Without bothering about the convergence, each element in $\hat{\mathcal{A}}_{\tilde\Theta}$ is a
formal power series in the $\hat{x}^\mu$'s for which the relation
$[\hat{x}^\mu, \hat{x}^\nu]=i\tilde{\Theta}^{\mu\nu}$ holds. The Moyal
algebra can be  also defined as  the linear space of smooth and rapidly decreasing
functions equipped with the  NC star product.
 The algebra of functions of such noncommuting coordinates can be represented
by the algebra of functions on ordinary spacetime, equipped with a noncommutative $\star-$product, i.e.
$(\hat{\mathcal{A}}_{\tilde{\Theta}}, .)\cong (\mathcal{A}_{\tilde{\Theta}},\star)$.
\, It is obvious that this deformation breaks the classical Lorentz symmetry. However, one can construct a deformation
of Lorentz symmetry such that it is a symmetry of relation $[x^\mu,x^\nu]_\star=i\widetilde{\Theta}^{\mu\nu}(x)$. The way
to do this construction is to use  a Lie algebra $\cal{G}$ generated by D elements $t^\mu$. The twist element
$\mathcal{F}\in \mathcal{U}_{\cal{G}}\otimes \mathcal{U}_{\cal{G}}$, where $\mathcal{U}_{\cal{G}}$ is the universal
enveloping algebra of $\cal{G}$. $\mathcal{U}_{\cal{G}}$ is a Hopf algebra and there is a linear map, called coproduct
$\Delta:\mathcal{U}_{\cal{G}}\rightarrow \mathcal{U}_{\cal{G}}\otimes \mathcal{U}_{\cal{G}}$ such that
$\Delta(t^\mu)=t^\mu\otimes 1+1\otimes t^\mu$. The main property that $\mathcal{F}$
has to satisfy is the cocycle condition $\mathcal{F}\otimes 1(\Delta\otimes id)\mathcal{F}=1\otimes
\mathcal{F}(id\otimes\Delta)\mathcal{F}$; it ensures the associativity of the star-product,  important
in the construction of this deformation. Aschieri et al \cite{aschieri} showed that, using the commuting
vector field $X_a=e_a^\mu\partial_\mu\in
T\mathbb{R}^D_{\tilde{\Theta}},$ i.e. $[X_a,X_b]=0$, the deformation of usual Moyal algebra remains associative, that 
is not the case  in  general. 

This work aims at deepening, by  using  an operator method, the investigations started in our recent paper  \cite{hd2}
  on the spectrum and  states of the harmonic
oscillator on twisted Moyal plane, where rather
 a Moyal-star-algebraic approach has been used and has produced a series of new unexpected results. 
As in the latter study,  we  deform the Moyal algebra with
noncommuting vectors fields $X_a,$ i.e. $[X_a,X_b]\neq 0$,  and suitable condition on the parameter $e_a^\mu$
 so that the star-product
obeys the associativity property. All the results
of our investigation obviously remain  valid  when the considered vector fields commute. We then
  solve the harmonic oscillator eigenvalue problem in  such a pertinent way to
 enjoyably produce  appropriate physical quantities, i.e. eigen energies and  states. 

This paper is organized as follows.
 In section 2,  we recall the  star-product properties used in the sequel, and deal 
with a comparative study of the harmonic oscillator in   both
 ordinary and twisted  Moyal planes. In section 3, the properties of harmonic oscillator
at infinity are given.
Section 4 is devoted to final remarks.

\section{Harmonic oscillator in twisted Moyal space}
\subsection{Brief review of useful twisted Moyal product properties}
As a matter of completeness, let us immediately provide a quick general survey 
of useful properties of twisted Moyal product which are used in the sequel.

In the  context of a dynamical noncommutative
field theory, the vector field can be generalized to take the
form $X_{a}=e_{a}^{\mu}(x)\partial_{\mu}$, where $e_{a}^{\mu}(x)$
is a tensor depending on the coordinate functions in the complex general  linear  matrix
group of order $D$ denoted by
GL$(D,\mathbb{C})$ \cite{aschieri}. The star product  takes the form
\begin{eqnarray}\label{pppp}
 (f\star g)(x) =  m\Big \{e^{i \frac{\Theta^{ab}}{2}
 X_{a}\otimes X_{b}} f(x)\otimes g(x)
 \Big \} \quad x\in \mathbb{R}_{\tilde\Theta}^D \quad  \forall f,g\in C^{\infty}(\mathbb{R}_{\tilde\Theta}^D)
 \end{eqnarray}
 and the vielbeins are given by the infinitesimal affine
transformation  as
\begin{eqnarray}\label{vie}
e_{a}^{\mu}(x)=\delta_{a}^{\mu}+\omega_{ab}^{\mu}x^b,
\end{eqnarray}
 where $\omega_{ab}\in \;$GL$(D,\mathbb{C})$ is skewsymmetric. Using (\ref{vie}), the  non vanishing
Lie bracket peculiar to  the non-coordinate base   \cite{hd2}
\begin{eqnarray}
[X_a,X_b]=e_\nu^c\Big[e_a^\mu\partial_\mu
e_b^\nu-e_b^\mu\partial_\mu e_a^\nu\Big] X_c= C_{ab}^c X_c
\end{eqnarray}
 is here simply reduced to
\begin{eqnarray}\label{rela}
[X_a,X_b]=\omega_{ba}^\mu\partial_{\mu}-\omega_{ab}^\mu\partial_{\mu}=-2\omega_{ab}^\mu\partial_{\mu}.
\end{eqnarray}
Besides, the dynamical star product  (\ref{pppp}) can be now expressed as
\begin{eqnarray}\label{prod}
(f\star g)(x)=m\Big[\exp\Big(\frac{i}{2}\theta
e^{-1}\epsilon^{\mu\nu}\partial_\mu\otimes\partial_\nu\Big)(f\otimes
g)(x)\Big]
\end{eqnarray}
where  $e^{-1}=:det(e_{a}^{\mu})=1+\omega_{12}^{1}x^{2}-\omega_{12}^{2}x^{1}$;\,
$\epsilon^{\mu\nu}$ is the symplectic tensor in two dimensions, $(D=2),$ with components
$\epsilon^{12}=-\epsilon^{21}=1,\,\,
\epsilon^{11}=\epsilon^{22}=0$.
The coordinate function commutation relation becomes
 $[x^\mu, x^{\nu}]_{\star}=i\widetilde{\Theta}^{\mu\nu}=i(\Theta^{\mu\nu}-\Theta^{a[\mu}\omega_{ab}^{\nu]}x^{b})$
 which can be reduced to the usual Moyal space relation, as expected, by setting
 $\omega_{ab}^{\mu}=[0]$.
One can  check that the Jacobi identity is also well satisfied, i.e.
 \begin{eqnarray}\label{id}
 [x^\mu,[x^\nu, x^{\rho}]_{\star}]_{\star}+[x^\rho,[x^\mu, x^\nu]_{\star}]_{\star}
 +[x^\nu,[x^\rho, x^\mu]_{\star}]_{\star}=\Theta^{b\mu}\Theta^{d[\nu}\omega_{bd}^{\rho]}=0
\end{eqnarray}
conferring a Lie algebra structure to the defined twisted Moyal
space.
This identity   ensures the  associativity of the
star-product (\ref{pppp}) and implies that
\begin{eqnarray}
 \widetilde{\Theta}^{\sigma\rho}\partial_{\rho}\widetilde{\Theta}^{\mu\nu}
 +\widetilde{\Theta}^{\nu\rho}\partial_{\rho}\widetilde{\Theta}^{\sigma\mu}
 +\widetilde{\Theta}^{\mu\rho}\partial_{\rho}\widetilde{\Theta}^{\nu\sigma}=0.
\end{eqnarray}
Therefore the algebra $\mathcal{A}_{\tilde{\Theta}}$ with this approach is the associative algebra and
its universal algebra $\mathcal{U}(\mathcal{A}_{\tilde{\Theta}})$ is a Hopf algebra. The twisted star-product
(\ref{pppp}) is then well defined.
Remark that with the relation
(\ref{rela}), the requirement that $\omega_{ab}$ is a symmetric
tensor trivially ensures  the associativity of the star product.
In the interesting particular case addressed in this work,  the associativity
of the star product (\ref{pppp}) is guaranteed even with the non symmetric
tensor $\omega_{ab}$. See proof in \cite{hd2}.
Besides, 
\begin{proposition}\label{p1.1}
 If $f$ and $g$ are two Schwartz functions on
 $\mathbb{R}^{2}_{\Theta}$,
then $f\star g$ is also a Schwartz function on $\mathbb{R}^{2}_{\Theta}$.
\end{proposition}
and
\begin{proposition}
 Notwithstanding  the condition  $[X_a,X_b]\neq 0,$ i.e.  $\omega_{ab}^\mu$ is skew-symmetric, the defined twisted
$\star-$product   remains noncommutative and associative.
\end{proposition}
{\bf Proof}: See  \cite{hd2}. $\square$

The tensor  $\widetilde{\Theta}^{\mu\nu}$ can be  decomposed in our case as:
\begin{eqnarray}
(\widetilde{\Theta})^{\mu\nu}=(\Theta)^{\mu\nu}-(\Theta^{a[\mu
}\omega_{ab}^{\nu]})x^b =\Big(\begin{array}{cc}
0& \theta e^{-1}\\
-\theta e^{-1}&0
\end{array}\Big)
\end{eqnarray}
and the  twisted Moyal
star-product satisfies the useful relation
\begin{eqnarray}\label{f}
x^{\mu}\star f=x^{\mu}f+\frac{i}{2}\Theta^{ab}e_{a}^{\mu}e_{b}^{\rho}\partial_{\rho}f
\;\mbox{ and }\,
f\star x^{\mu}=x^{\mu}f-\frac{i}{2}\Theta^{ab}e_{a}^{\mu}e_{b}^{\rho}\partial_{\rho}f.
\end{eqnarray}
The anticommutator and commutator star brackets  of $x^{\mu}$ and
$f$ can be immediately deduced as follows:
\begin{eqnarray}
\{x^{\mu}, f\}_{\star}=2x^{\mu}f,\quad [x^{\mu}, f]_{\star}=i\Theta^{ab}e_{a}^{\mu}e_{b}^{\rho}\partial_{\rho}f.
\end{eqnarray}
The relations (\ref{f}) can be detailed for $x^{\mu},\; \mu=1, 2$
as:
\begin{eqnarray}\label{11}
x^{1}\star f=x^{1}f+\frac{i}{2}\theta e^{-1}\partial_{2}f\quad\quad
f\star x^{1}=x^{1}f-\frac{i}{2}\theta e^{-1}\partial_{2}f
\end{eqnarray}
\begin{eqnarray}\label{12}
x^{2}\star f=x^{2}f-\frac{i}{2}\theta e^{-1}\partial_{1}f\quad\quad
f\star x^{2}=x^{2}f+\frac{i}{2}\theta e^{-1}\partial_{1}f
\end{eqnarray}
giving rise to the  creation and annihilation functions
\begin{eqnarray}
a=\frac{x^1 +ix^2}{\sqrt{2}}\quad\quad\quad \bar{a}=\frac{x^1-ix^2}{\sqrt{2}}
\end{eqnarray}
 with the commutation relation
$[a,\bar{a}]_{\star}=\theta e^{-1}$.
It then becomes a matter of algebra to use the  transformations of the vector fields $\partial_{1}$ and $\partial_{2}$
into $\partial_{a}=:\frac{\partial}{\partial a}$ and $\partial_{\bar{a}}=:\frac{\partial}{\partial \bar{a}}$ and vice-versa to infer
$e^{-1}=1-a\omega-\bar{a}\bar{\omega}$ and  $e= 1+a\omega+\bar{a}\bar{\omega}$,
where
\begin{eqnarray}
\omega=:\frac{\omega_{12}^{2}+i\omega_{12}^{1}}{\sqrt{2}}\quad\mbox{ and }\quad \bar{\omega}=:
\frac{\omega_{12}^{2}-i\omega_{12}^{1}}{\sqrt{2}}.
\end{eqnarray}
There result the  useful relations
\begin{eqnarray}
\frac{\partial e^{-1}}{\partial a}=-\omega,\quad \frac{\partial e^{-1}}{\partial \bar{a}}=-\bar{\omega}\mbox{ and for }
 k\in \mathbb{Z},\quad \omega e^{k}=\omega,
\quad \bar{\omega}e^{k}=\bar{\omega}.
\end{eqnarray}
Expressing the twisted $\star-$product (\ref{prod}) in terms of vector fields $\partial_{a}$ and $\partial_{\bar{a}}$  as
\begin{eqnarray}\label{starstar}
(f\star g)(a,\bar{a})&=&{\rm m}\Big[\sum_{n=0}^{\infty}\sum_{k=0}^{n}
\frac{(-1)^{n-k}}{k!(n-k)!}(\frac{1}{2}\theta e^{-1})^{n}\cr
&&\times (\partial_{a}\otimes\partial_{\bar{a}})^{k}(\partial_{\bar{a}}\otimes\partial_{a})^{n-k}(f\otimes g)(a,\bar{a})\Big]
\end{eqnarray}
and using equations (\ref{11}) and (\ref{12}) (or independently (\ref{starstar})) yield
\begin{eqnarray}
a\star f=\Big(a+\frac{\theta e^{-1}}{2}\frac{\partial}{\partial \bar{a}}\Big)f \quad
\quad \bar{a}\star f=\Big(\bar{a}-\frac{\theta e^{-1}}{2}\frac{\partial}{\partial a}\Big)f
\end{eqnarray}
\begin{eqnarray}
f\star a=\Big(a-\frac{\theta e^{-1}}{2}\frac{\partial}{\partial \bar{a}}\Big)f\quad\quad f\star \bar{a}=\Big(\bar{a}+\frac{\theta e^{-1}}{2}\frac{\partial}{\partial a}\Big)f.
\end{eqnarray}

\subsection{Physical states and spectrum}
From the above derived results, the  twisted ho Hamiltonian operator $H=a\bar{a}$ is defined by the relation
\begin{eqnarray}
H\star(.)
&=&\frac{1}{2}\Big[(x^1)^2+(x^2)^2 + \Big(i\theta e^{-1}x^1-\frac{\theta^2}{4}\omega_{12}^1\Big)\partial_2\cr
&& -\Big(i\theta e^{-1}x^2-\frac{\theta^2}{4}\omega_{12}^2\Big)\partial_1
- \frac{\theta^2}{4}e^{-2}(\partial_1^2 + \partial_2^2)\Big]\equiv \frac{1}{2}\mu_1
\end{eqnarray}
with the domain
\begin{eqnarray}
 \mathcal{D}(H\star)= \left\{f\in L^{2}(\mathbb{R}_{\Theta}^{2})\mid f, f_{x^1}, f_{x^2} \in \mathcal{AC}_{loc}(\mathbb{R}_{\Theta}^{2});
 \frac{\mu_1}{2}f\in  L^{2}(\mathbb{R}_{\Theta}^{2})\right\}.
\end{eqnarray}
 $\mathcal{AC}_{loc}(\mathbb{R}_{\Theta}^{2})$
 denotes the set of  locally absolutely continuous functions on $\mathbb{R}_{\Theta}^{2}$. Similarly,

\begin{eqnarray}
(.)\star H
& = &\frac{1}{2}\Big[(x^1)^2+(x^2)^2 - \Big(i\theta e^{-1}x^1-\frac{\theta^2}{4}\omega_{12}^1\Big)\partial_2\cr
&+&\Big(i\theta e^{-1}x^2-\frac{\theta^2}{4}\omega_{12}^2\Big)\partial_1
- \frac{\theta^2}{4}e^{-2}(\partial_1^2 + \partial_2^2)\Big]\equiv \frac{1}{2}\mu_2
\end{eqnarray}
defined in the domain
\begin{eqnarray}
 \mathcal{D}(\star H)= \left\{f\in L^{2}(\mathbb{R}_{\Theta}^{2})\mid f, f_{x^1}, f_{x^2} \in \mathcal{AC}_{loc}(\mathbb{R}_{\Theta}^{2});
 \frac{\mu_2}{2}f\in  L^{2}(\mathbb{R}_{\Theta}^{2})\right\}.
\end{eqnarray}
Setting
$\omega_2=\omega_{12}^1$, $\omega_1=\omega_{12}^2$, $e^{-1}=1+\omega_2x_2-\omega_1x_1,$
 the eigenvalue equation can be written
as
\begin{eqnarray}\label{equ1}
 \Big[x_1^2+x_2^2-\frac{\theta^2\omega_2}{4}\frac{\partial}{\partial x_2}+
\frac{\theta^2\omega_1}{4}\frac{\partial}{\partial x_1}-\frac{\theta^2 e^{-2}}{4}\Big(\frac{\partial^2}{\partial x_1^2}+
\frac{\partial^2}{\partial x_2^2}\Big)\Big]f=2Ef
\end{eqnarray}
with
\begin{eqnarray}\label{equ2}
 \Big(x_1\frac{\partial}{\partial x_2}-x_2\frac{\partial}{\partial x_1}\Big)f=0
\end{eqnarray}
which naturally suggests the use of  polar coordinates as follows:
\begin{eqnarray}\label{28noi}
x_1=r\cos\alpha,\qquad x_2=r\sin\alpha,\quad x_1^2+x_2^2=r^2.
\end{eqnarray}
The variable change (\ref{28noi}) transforms the equation (\ref{equ1}) into the form:
\begin{eqnarray}\label{equ4}
&& \Big[r^2-\frac{\theta^2\omega_2}{4}\Big(\sin\alpha\frac{\partial}{\partial r}+
\frac{\cos\alpha}{r}\frac{\partial}{\partial \alpha}\Big)+
\frac{\theta^2\omega_1}{4}\Big(\cos\alpha \frac{\partial}{\partial r}-\frac{\sin\alpha}{r}\frac{\partial}{\partial \alpha}
\Big)\cr
&&-\frac{\theta^2 }{4}(1+2\omega_2 r\sin\alpha-2\omega_1 r\cos\alpha)\Big(\frac{\partial^2}{\partial r^2}+
\frac{1}{r}\frac{\partial}{\partial r}+\frac{1}{r^2}\frac{\partial^2}{\partial \alpha^2}\Big)\Big]f=2Ef.
\end{eqnarray}
Separating the variables in the function $f$ as follows:
\begin{eqnarray}\label{equ5}
 f(r,\alpha)=\chi(r)g(\alpha)\, \mbox{ where }\, g(\alpha)=e^{ik\alpha},
\end{eqnarray}
and replacing the result   in (\ref{equ4}), we get
\begin{eqnarray}\label{equ6}
&&\Big[r^2-2E -\frac{\theta^2\omega_2}{4}\Big(\sin\alpha\frac{\partial}{\partial r}+
ik\frac{\cos\alpha}{r}\Big)+
\frac{\theta^2\omega_1}{4}\Big(\cos\alpha \frac{\partial}{\partial r}-ik\frac{\sin\alpha}{r}\Big)\cr
&&-\frac{\theta^2 }{4}(1+2\omega_2 r\sin\alpha-2\omega_1 r\cos\alpha)\Big(\frac{\partial^2}{\partial r^2}+
\frac{1}{r}\frac{\partial}{\partial r}-\frac{k^2}{r^2}\Big)\Big]\chi(r)=0
\end{eqnarray}
resulting in the unique equation
\begin{eqnarray}\label{equ7}
&&\Big[r^2-2E -\frac{\theta^2\omega_2}{4}\sin\alpha\frac{\partial}{\partial r}+
\frac{\theta^2\omega_1}{4}\cos\alpha \frac{\partial}{\partial r}\cr
&&-\frac{\theta^2 }{4}(1+2\omega_2 r\sin\alpha-2\omega_1 r\cos\alpha)\Big(\frac{\partial^2}{\partial r^2}+
\frac{1}{r}\frac{\partial}{\partial r}-\frac{k^2}{r^2}\Big)\Big]\chi(r)=0
\end{eqnarray}
with the constraint relation
\begin{eqnarray}\label{equ8}
\omega_1\sin\alpha +\omega_2\cos\alpha=0.
\end{eqnarray}
 Finally,  introducing the latter in the  equation (\ref{equ7}) generates the appropriate differential equation:
\begin{eqnarray}\label{equ10}
\Big[r^2-2E-\frac{\theta^2}{4}\Big(\frac{\partial^2}{\partial r^2}+
\frac{1}{r}\frac{\partial}{\partial r}-\frac{k^2}{r^2}\Big) +\frac{\theta^2\omega_1}{4\cos\alpha}
\Big(2r\frac{\partial^2}{\partial r^2}+
3\frac{\partial}{\partial r}-\frac{2k^2}{r}\Big)\Big]\chi(r)=0. 
\end{eqnarray}
As a matter of result comparison, let us now search for the solutions of this equation 
by considering both ordinary and twisted Moyal spaces.
\begin{itemize}
\item[({\bf B1})] Case of the ordinary Moyal space

It corresponds to $\omega_1=0$ reducing the equation (\ref{equ10})  to
\begin{eqnarray}\label{equ11'}
\Big[\frac{\partial^2}{\partial r^2}+
\frac{1}{r}\frac{\partial}{\partial r}-\frac{k^2}{r^2}-\frac{4}{\theta^2}(r^2-2E)\Big]\chi(r)=0.
\end{eqnarray}
Making the variable change $r^2-E=u\Rightarrow 2rdr=du,$ 
\begin{eqnarray}
r\frac{\partial}{\partial r}=2(u+E) \frac{\partial}{\partial u},\,\quad\, \frac{\partial^2}{\partial r^2}
=4(u+E)\frac{\partial^2}{\partial u^2} +2\frac{\partial}{\partial u}
\end{eqnarray}
and the equation (\ref{equ11'}) takes the form
\begin{eqnarray}
\Big[4(u+E)^2\frac{\partial^2}{\partial u^2}+4(u+E)\frac{\partial}{\partial u}
-\frac{4}{\theta^2}(u^2-E^2+\frac{\theta^2 k^2}{4})\Big]\chi(u)= 0.
\end{eqnarray}
Then, choosing $u+E=\rho$ further simplifies the expressions to give
\begin{eqnarray}\label{sing}
\Big[\rho^2 \frac{\partial^2}{\partial\rho^2 } +\rho\frac{\partial}{\partial \rho}-\frac{1}{\theta^2}\Big(\rho^2-2E\rho+
\frac{\theta^2 k^2}{4}\Big)\Big]\chi(\rho)=0
\end{eqnarray}
with singularities at $\rho=0$ and at $\rho=\infty$. In the vicinity of $\rho=0,$ we find $\chi(\rho)$
 proportional to $\rho^\nu,$  and at infinity  $\chi(\rho)=e^{ -B\rho}$. Therefore,
if we write the solution as follows:
\begin{eqnarray}
\chi(\rho)=\rho^\nu e^{-B\rho}F(\rho),\mbox{ with }\,\nu=1\pm\sqrt{1+\frac{k^2}{4}},\,\, B=\frac{1}{\theta}
\end{eqnarray}
the resulting differential equation for $F(\rho)$  turns out to be in the form
\begin{eqnarray}\label{new2011}
\Big[\rho \frac{\partial^2}{\partial\rho^2 } +(2\nu+1-2B\rho)\frac{\partial}{\partial \rho}+\Big(\frac{2E}{\theta^2}
-2\nu B-B\Big)\Big]F(\rho)=0
\end{eqnarray}
which, with
\begin{eqnarray}
2B\rho=\rho',\,\,  a=\frac{1}{2B}\Big(\frac{2E}{\theta^2}
-2\nu B-B\Big),\,\, b=2\nu+1,
\end{eqnarray}
is transformed into the Kummer confluent hypergeometric equation
\begin{eqnarray}
\Big[\rho' \frac{\partial^2}{\partial\rho^{'2} } +(b-\rho')\frac{\partial}{\partial \rho'}+a\Big]F(\rho')=0
\end{eqnarray}
whose  the general solution  is given by
\begin{eqnarray}
F(\rho')=A_1\Phi(a,b;\rho')+A_2\rho^{'1-b}\Phi(a-b+1,2-c;\rho'),\,\, A_1,\,A_2\in \mathbb{R},
\end{eqnarray}
where
\begin{eqnarray}
\Phi(a,b;\rho')=\sum_{n=0}^\infty\frac{(a)_n}{(b)_n n!}\rho^{'n},\,\,\, (a)_n=\frac{\Gamma(a+n)}{\Gamma(a)}=a(a+1)\cdots
(a+n-1).
\end{eqnarray}
The solution of (\ref{sing}) then becomes
\begin{eqnarray}\label{sol1}
\chi(\rho)= \rho^\nu e^{-B\rho}\Big[A_1\Phi(a,b;2B\rho)+A_2(2B \rho)^{1-b}\Phi(a-b+1,2-c;2B\rho)
 \Big],
\end{eqnarray}
where
\begin{eqnarray}
\nu=1\pm\sqrt{1+\frac{k^2}{4}},\,\, B=\frac{1}{\theta},\,\, 2B\rho=\rho',\,\,
 a=\frac{E}{\theta}
-\nu-\frac{1}{2},\,\, b=2\nu+1.
\end{eqnarray}
The physical states i.e. bounded states occur only for $B>0$, i.e. $\theta>0,$ since the confluent series $\Phi(a,b;\rho),$
for large values of $\rho,$ is proportional to $e^\rho$ so that $\chi$ diverges for $\rho\rightarrow\infty$. Further, the 
second term of (\ref{sol1}) has a regular singularity at $\rho\rightarrow 0$ if $b>1$. Hence the physical states 
(see details in the case ({\bf B2}) below) are given by
\begin{eqnarray}\label{solprof}
  f(r,\alpha)=\sqrt{\frac{2^{2\nu_p+\frac{3}{2}} B^{2\nu_p+\frac{3}{2}}
[(2\nu_p+1)_p]^2\Gamma(2\nu_p+1)}{\pi p!\Gamma(2\nu_p+p+1)
\Gamma(2\nu_p+\frac{3}{2})}}. r^{2\nu} e^{-B r^2}\Phi(a,b;2Br^2)e^{i\alpha k},
\end{eqnarray}
corresponding to the eigen-energies
\begin{eqnarray}\label{energy}
E_{l,k}^{\pm}=\theta\Big(\frac{3}{2}\pm\sqrt{1+\frac{k^2}{4}} -l\Big),
\end{eqnarray}
with $a=-l,\,\, l=0,1,2,\cdots$.
Finally, in accordance with  \cite{Gracia-Bondia} (and references therein),  $k$ must 
 satisfy the relation
\begin{eqnarray}\label{condit}
k=\pm \sqrt{(n+l-1)^2-4},\mbox{ where } n \in\mathbb{N}\Rightarrow E_{l,\nu}^{+}= E_n= \theta\Big(n+\frac{1}{2}\Big).
\end{eqnarray}
The solution (\ref{solprof}) can be transformed into Laguerre 
or Hermite polynomials by using a suitable transformation. Further 
 by adopting an appropriate normalization
constant, one can show that it is well equivalent to  the result  given in  \cite{Gracia-Bondia}. See  Appendix.

\item[({\bf B2})] Case of the twisted Moyal space

Consider now equation (\ref{equ10}) in the case when $\omega_1\neq 0$. This equation can be re-expressed as follows:
\begin{eqnarray}\label{total}
 \Big[r^2-2E+\frac{\theta^2 k^2}{4r^2}\Big(1-\frac{2\omega_1 r}{\cos\alpha}\Big)-
\frac{\theta^2}{4r}\Big(1-\frac{3\omega_1 r}{\cos\alpha}\Big)\frac{\partial}{\partial r}
-\frac{\theta^2}{4}\Big(1-\frac{2\omega_1 r}{\cos\alpha}\Big)\frac{\partial^2}{\partial r^2}\Big]\chi(r)=0.\nonumber\\
\end{eqnarray}
As $\omega_1 $ is an infinitesimal parameter, the equation (\ref{total}) can be reduced to
\begin{eqnarray}\label{new1}
 \Big[\frac{\partial^2}{\partial r^2}+\frac{1}{r}\Big(1-\frac{\omega_1 r}{\cos\alpha}\Big)\frac{\partial}{\partial r}-
\frac{k^2}{r^2}-\frac{4}{\theta^2}\Big(1+\frac{2\omega_1 r}{\cos\alpha}\Big)(r^2-2E)\Big]\chi(r)=0.
\end{eqnarray}
Defining $\chi$ in terms of series, $\chi(r)=\sum_n a_n r^n,$ with $a_n$  such that
$a_{2p+1}=0, \forall p\in \mathbb{N},$ then (\ref{new1}) is re-expressed as
\begin{eqnarray}
&&\sum_{n}(n^2-k^2)a_n r^{n-2}-\frac{\omega_1}{\cos\alpha}\sum_{n}n a_n r^{n-1}+\frac{8E}{\theta^2}\sum_{n} a_n r^n
\cr
&&+\frac{16E\omega_1}{\theta^2\cos\alpha}\sum_n a_n r^{n+1}-\frac{4}{\theta^2} \sum_n a_n r^{n+2}-\frac{8\omega_1}{\theta^2\cos\alpha}\sum_n a_n r^{n+3}=0.
\end{eqnarray}
The sequence $a_n$ satisfies the recurrence relation
\begin{eqnarray}\label{rel-new}
&&\Big((n+2)^2-k^2\Big)a_{n+2}- \frac{\omega_1}{\cos\alpha}(n+1)a_{n+1}+\frac{8E}{\theta^2} a_n+
\frac{16E\omega_1}{\theta^2\cos\alpha}a_{n-1}\cr
&&-\frac{4}{\theta^2}a_{n-2}-\frac{8\omega_1}{\theta^2\cos\alpha}a_{n-3}=0.
\end{eqnarray}
If $n=0,$  then $a_2=-\frac{8E}{\theta^2}\frac{1}{4-k^2}a_0$. Similarly,
if $n=1,$ $a_2=\frac{8E}{\theta^2}a_0$ and we infer that $k=\pm\sqrt{5}$. If $n=2$ and $n=3$, then we obtain the values
\begin{eqnarray}
&&a_4=\Big[\Big(-\frac{8E}{\theta^2}\Big)^2\frac{1}{2^2-k^2}.\frac{1}{4^2-k^2}
+\frac{4}{\theta^2}
\frac{1}{4^2-k^2}\Big]a_0, \;\;\mbox{and}\cr
&&
a_4=\Big[\Big(-\frac{16E}{\theta^2}\Big)^2\frac{1}{2}.\frac{1}{4}
-\frac{8}{\theta^2}
\frac{1}{4}\Big]a_0,
\end{eqnarray}
respectively,
 implying  $k=3\sqrt{2}.$ Continuing this procedure, we succeed in
 separating the relation (\ref{rel-new}) into new recurrence relations
\begin{eqnarray}\label{recc1}
\Big((n+2)^2-k^2\Big)a_{n+2} +\frac{8E}{\theta^2} a_n-\frac{4}{\theta^2}a_{n-2}=0
\end{eqnarray}
 and
\begin{eqnarray}\label{recc2}
 - (n+2)a_{n+2}+
\frac{16E}{\theta^2}a_{n}-\frac{8}{\theta^2}a_{n-2}=0.
\end{eqnarray}
which 
are equivalent if and only if
\begin{eqnarray}
k=\pm \sqrt{\frac{(n+2)(2n+5)}{2}}.
\end{eqnarray}
As $n$ is an even integer in the equations (\ref{recc1}) and (\ref{recc2}),  we obtain
\begin{eqnarray}\label{kep}
k=k_p=\pm\sqrt{(p+1)(4p+5)}, \,\quad\mbox{ where }\, \quad \frac{n}{2}=p\in\mathbb{N}.
\end{eqnarray}
\begin{remark}
The recurrence relation (\ref{recc1})  can be also deduced from the case ({\bf B1})  by solving, with the same series solution method, 
the equation
(\ref{equ10}) for $\omega_1=0.$ 
\end{remark}


The Hilbert space structure on $ \mathcal{S}(\mathbb{R}^2_{\tilde{\Theta}})$ is defined by
the scalar product
\begin{eqnarray}
<f,g>&=&:\int_{-\infty}^{+\infty}\,\int_{-\infty}^{+\infty}\,\,dx_1dx_2\,\,\overline{f(x_1,x_2)}g(x_1, x_2),\quad 
f,g\in \mathcal{S}(\mathbb{R}^2_{\tilde{\Theta}})\cr
&=&\int_0^{2\pi}\,d\alpha\,\int_{0}^{+\infty}\, r^2dr\,\overline{f(r,\alpha)}g(r, \alpha).
\end{eqnarray}
There results the normalization condition
\begin{eqnarray}
\int_0^{2\pi}\,d\alpha\,\int_{0}^\infty\, r^2dr\,  \overline{f_{p}(r,\alpha)}f_{q}(r,\alpha)=\delta_{pq}. 
\end{eqnarray}
Therefore,  the following result is in order.
\begin{proposition}
The normalised eigenstates and eigenenergies of the harmonic oscillator on twisted Moyal plane are given, respectively, by
\begin{eqnarray}\label{solutionfinal}
f_{p}(r,\alpha)= \sqrt{\frac{2^{2\nu_p+\frac{3}{2}} B^{2\nu_p+\frac{3}{2}}
[(2\nu_p+1)_p]^2\Gamma(2\nu_p+1)}{\pi p!\Gamma(2\nu_p+p+1)
\Gamma(2\nu_p+\frac{3}{2})}} r^{2\nu_p} e^{-B r^2}\Phi(-p,2\nu_p+1;2B r^2) e^{ i\alpha k_p}\nonumber\\
\end{eqnarray}
and
\begin{eqnarray}\label{low}
&&E_{p}^{(\pm)}=\theta\Bigg(\frac{3}{2}\pm\sqrt{1+\frac{(p+1)(4p+5)}{4}}-p\Bigg), \quad p\in \mathbb{N}
\end{eqnarray}
with
\begin{eqnarray}
 \nu_p=1\pm\sqrt{1+\frac{k_p^2}{4}},\,\, B=\frac{1}{\theta},\,\,
 a=\frac{E_p}{\theta}
-\nu_p-\frac{1}{2}=:-p,\,\, b=2\nu_p+1.
\end{eqnarray}
\end{proposition}
{\bf Proof:}
Using the relation (\ref{Lag}) given in Appendix, the identity
\begin{eqnarray}\label{magic}
\int_{0}^{+\infty}\, dz\, e^{-z}z^{\sigma+\delta} L_n^\sigma(z) L_m^\sigma(z)=\delta_{nm}
\frac{\Gamma(n+\sigma+1)\Gamma(\sigma+
\delta+1)}{n!\Gamma(\sigma+1)} 
\end{eqnarray}
and (\ref{solutionfinal}), we arrive at the normalization condition
\begin{eqnarray}\label{relanew}
&&\int_0^{2\pi}\,d\alpha\,\int_{0}^{+\infty}\, r^2dr\,  \overline{f_{p}(r,\alpha)}f_{q}(r,\alpha)=|A_1|^2
\frac{2\pi}{2\sqrt{2}}\Big(\frac{1}{2}\Big)^{2\nu_p+1}\cr
&&\times B^{-\frac{3}{2}-2\nu_p}\frac{(p)!^2}{[(2\nu_p+1)_p]^2}\frac{\Gamma(p+2\nu_p+1)\Gamma(2\nu_p+
\frac{3}{2})}{p!\Gamma(2\nu_p+1)}
\delta_{pq}=\delta_{pq},
\end{eqnarray}
yielding 
\begin{eqnarray}
A_1^2=\frac{2^{2\nu_p+\frac{3}{2}} B^{2\nu_p+\frac{3}{2}}[(2\nu_p+1)_p]^2\Gamma(2\nu_p+1)}{\pi\Gamma(2\nu_p+p+1)
\Gamma(2\nu_p+\frac{3}{2})p!},
\end{eqnarray} 
where $\Gamma(.)$ is the gamma function defined by the relation $\Gamma(z)=:\int_0^\infty\, dt e^{-t}t^{z-1}$ while 
$(2\nu_p+1)_p$ is provided by the expression (\ref{topnew}) found in  Appendix. $\square$

Finally, we  conclude that
 the states of the harmonic oscillator in the ordinary Moyal plane given
by (\ref{sol1}) are similar to those
of the twisted Moyal space by replacing $k$ by $k_p$ found in (\ref{kep}). 
Figures $1$ and $2$ illustrate the energy spectrum behaviour versus $p$. 
 $E_p^{(+)}$ decreases from $3$ to its asymptotic value $1.5$ as $p$ increases. $E_p^{(-)}$ admits a lower limit
$0$ for $p=0$ and linearily varies as $-p$ with increasing values of $p$. Therefore one can conclude that  $E_p^{(+)}$
and  $E_p^{(-)}$ represent the scattering and bound state energies, respectively, of the harmonic oscillator in twisted Moyal plane. 
This is a novel feature that has not been observed in our previous investigation of the harmonic oscillator in twisted 
Moyal plane (\cite{hd2} and references therein).
\end{itemize}

\begin{figure}[htbp]
\begin{center}
\includegraphics[width=15cm]{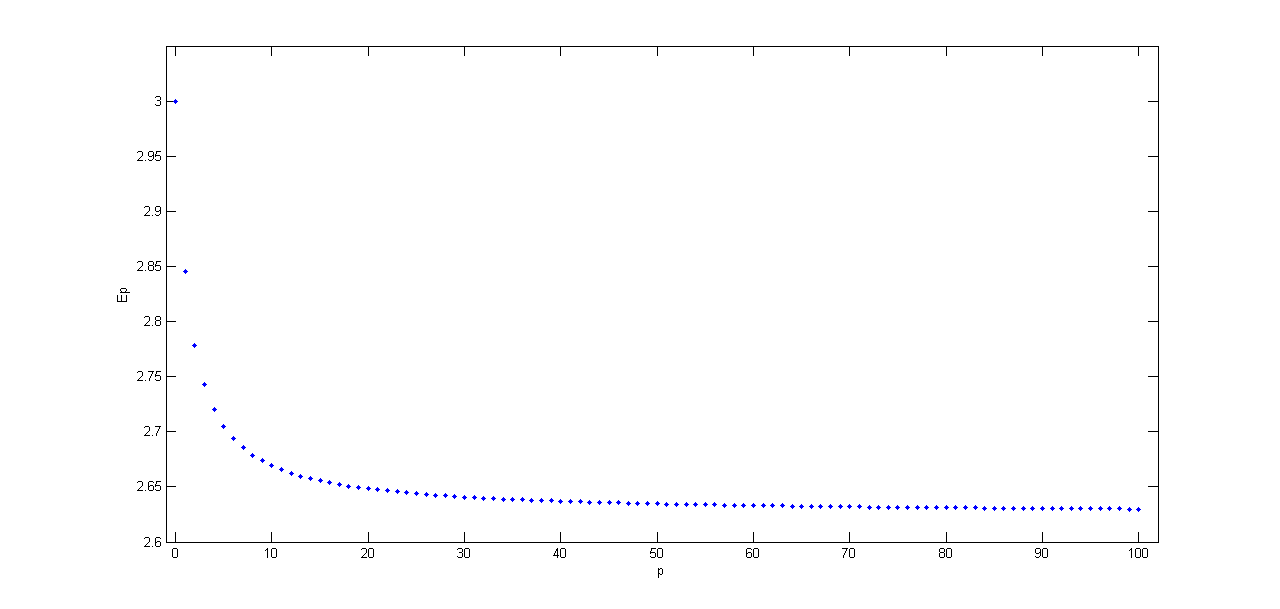}  
\end{center}
\caption{Energy  $E_p^{(+)}$ versus $p$ for $\theta=1.$}
\end{figure}
\begin{figure}[htbp]
 \begin{center}
\includegraphics[width=15cm]{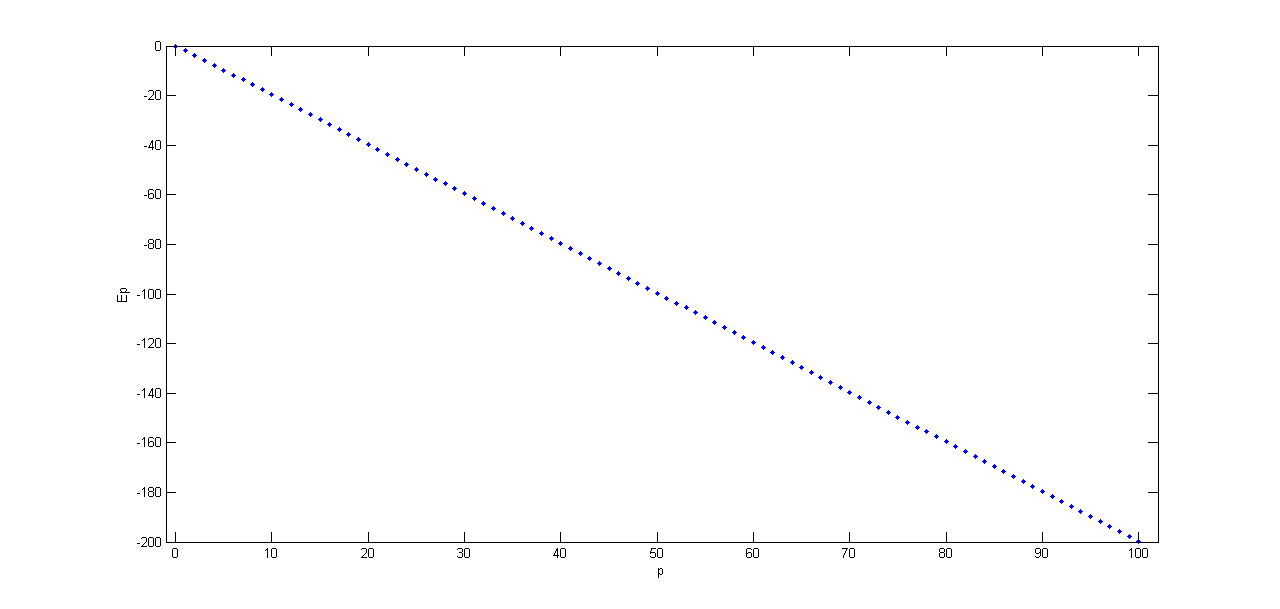} 
\end{center}
\caption{Energy $E_p^{(-)}$ versus $p$ for $\theta=1.$}
\end{figure}
\section{ Physical properties at infinity}
Equation (\ref{sing}) is reduced to
\begin{eqnarray}\label{zip}
\Big[\frac{\partial^2}{\partial\rho^2 } +\frac{1}{\rho}\frac{\partial}{\partial \rho}-\frac{1}{\theta^2}\Big(1-\frac{2E}{\rho}+
\frac{\gamma}{\rho^2}\Big)\Big]\chi(\rho)=0
\end{eqnarray}
where $\gamma=\frac{\theta^2 k^2}{4}$ and $\rho=r^2$. The potential $V(\rho)=-\frac{1}{\theta^2}\Big(1-\frac{2E}{\rho}+
\frac{\gamma}{\rho^2}\Big)\rightarrow -\frac{1}{\theta^2}$ if $\rho\rightarrow\infty$. In this limit, the equation
(\ref{zip}) is written as
\begin{eqnarray}\label{zipinfty}
\Big[\frac{\partial^2}{\partial\rho^2 } +\frac{1}{\rho}\frac{\partial}{\partial \rho}-B^2\Big]\chi(\rho)=0
\end{eqnarray}
where $B=\frac{1}{\theta}$. Suppose that $\chi(\rho\rightarrow\infty)=\sum_l a_l \rho^l$. Then (\ref{zipinfty}) is re-expressed as
\begin{eqnarray}
\sum_l l^2 a_l \rho^{l-2}-B^2\sum_l a_l \rho^l=0
\end{eqnarray}
and the series $a_l$ satisfies the recurrence relation
\begin{eqnarray}
 (n+2)^2a_{n+2}-B^2 a_n=0\Rightarrow a_{n+2}=\frac{B^2}{(n+2)^2}a_n,
\end{eqnarray}
with
\begin{eqnarray}
 a_n=\frac{B^2}{n^2}a_{n-2}=\frac{(B^2)^2}{n^2(n-2)^2}a_{n-4}=\cdots=\frac{(B^2)^l}{n^2(n-2)^2\cdots(n-2l+2)^2}a_{n-2l}.
\end{eqnarray}
If $l=\frac{n}{2}$ implying that n is an even integer, we get
\begin{eqnarray}
a_n&=&\frac{B^n}{n^2(n-2)^2\cdots 2^2} a_0=\Big[\frac{B^{n/2}}{n(n-2)\cdots 2}\Big]^2a_0\cr
&=&\Big[\frac{B^{n/2}}{2^{n/2} (n/2)(n/2-1)\cdots 1}\Big]^2a_0=
\Big[\frac{B^{n/2}}{2^{n/2} (n/2)!}\Big]^2a_0.
\end{eqnarray}
We then obtain
\begin{eqnarray}
a_{2n}=\frac{B^{2n}}{2^{2n}(n!)^2}a_0 \mbox{ and } \chi(\rho)_\infty=a_0\sum_{n=0}^\infty \frac{B^{2n}}{2^{2n}(n!)^2}\rho^{2n}.
\end{eqnarray}
Finally there results the following solution:
\begin{eqnarray}\label{jolie}
\chi(\rho)=a_0 e^{-\lambda\rho}\sum_{n} \frac{B^{2n}}{2^{2n}(n!)^2}\rho^{2n}\quad \lambda>0.
\end{eqnarray}
 Putting (\ref{jolie}) into the equation (\ref{zip}) yields
\begin{eqnarray}\label{74}
&(\lambda^2-B^2)\sum_n \frac{B^{2n}}{2^{2n}(n!)^2}\rho^{2n}+(2EB^2-\lambda)\sum_n\frac{B^{2n}}{2^{2n}(n!)^2}\rho^{2n-1}\cr
&-2\lambda\sum_n \frac{2n B^{2n}}{2^{2n}(n!)^2}\rho^{2n-1}-B^2\gamma\sum_n \frac{B^{2n}}{2^{2n}(n!)^2}\rho^{2n-2}\cr
&+\sum_n \frac{2nB^{2n}}{2^{2n}(n!)^2}\rho^{2n-2}+\sum_n \frac{2n(2n-1)B^{2n}}{2^{2n}(n!)^2}\rho^{2n-2}=0.
\end{eqnarray}
The parameter $\lambda$ satisfies the equation
\begin{eqnarray}
\lambda^2-\frac{3B}{\sqrt{\pi}}\lambda +\frac{2E_{0,k}B^3}{\sqrt{\pi}}-\frac{B^4\gamma}{4}=0
\end{eqnarray}
affording the solution
\begin{eqnarray}
\lambda=\frac{3B}{2\sqrt{\pi}}+\frac{1}{2}\sqrt{\frac{9B^2}{\pi}-\frac{8E_{0,k}B^3}{\sqrt{\pi}}+\gamma B^4}.
\end{eqnarray}
This relation bounds the energy by
\begin{eqnarray}
E_{0,k}\leq\frac{\sqrt{\pi}}{8}\Big(\frac{9}{\pi B}+\gamma B\Big)= \frac{\sqrt{\pi }\theta}{8}\Big(\frac{9}{\pi}+\frac{k^2}{4}\Big).
\end{eqnarray}
 Remark that, for $\rho\rightarrow \infty$, $\lambda=0$ and  the energy spectrum of the  ground state takes the form
\begin{eqnarray}
E_{0,k}^\infty=\frac{\gamma B\sqrt{\pi}}{8}=:\frac{k^2\theta\sqrt{\pi}}{32}.
\end{eqnarray}
We finally arrive at the following main result:
\begin{proposition}
The state of the harmonic oscillator in twisted Moyal space is given by
\begin{eqnarray}\label{79}
&f(r,\alpha)=a_0 e^{-\lambda r^2}\sum_{n} \frac{B^{n}}{2^{n}((n/2)!)^2}r^{2n} e^{ik\alpha},\quad
a_0\in\mathbb{R},\quad\alpha\in\mathbb{R}, \quad\gamma=\frac{\theta^2 k^2}{4},
\cr
& B=\frac{1}{\theta},\quad\lambda=\frac{3B}{2\sqrt{\pi}}+\frac{1}{2}\sqrt{\frac{9B^2}{\pi}-\frac{8E_{0,k}B^3}{\sqrt{\pi}}+\gamma B^4}
\end{eqnarray}
with the corresponding energy
\begin{eqnarray}\label{81}
E_{n,k}=\frac{(4n+3)\lambda}{2B^2}-\frac{(n+\frac{1}{2})!^2\lambda^2}{(n!)^2B^3}+\frac{(n+\frac{1}{2})!^2\gamma B}{4(n+1)!^2},
\,\,  E_{n,k}^\infty=\frac{(n+\frac{1}{2})!^2\gamma B}{4(n+1)!^2}.
\end{eqnarray}
\end{proposition}
{\bf Proof:} Equation \eqref{79} is immediately obtained by substitution of  \eqref{jolie} in \eqref{equ5}. The expression
\eqref{81} is the solution of \eqref{74} reducible, after some algebra, to
\begin{eqnarray}
 \frac{B^3(n!)^2E_{n,k}}{(n+\frac{1}{2})!^2}+\lambda^2-
\frac{(4n+3)(n!)^2\lambda B}{2(n+\frac{1}{2})!^2} -\frac{\gamma B^4}{4(n+1)^2}=0.\square
\end{eqnarray}
\section{Final remarks}
By deepening the analysis of the physical properties of the harmonic
oscillator on twisted Moyal plane, this work has proved that one can retrieve useful physical quantities, 
i.e.  physical states with real energies even in such deformed situation, thanks to the
 efficiency of the  used operator approach. 
 The twisted Moyal space favours the appearence of both seattering and bound states for the particle subject to a
harmonic potential. 

\section*{Acknowledgements}
This work is partially supported by the ICTP through the
OEA-ICMPA-Prj-15. The ICMPA is in partnership with the Daniel
Iagolnitzer Foundation (DIF), France.

\section*{Appendix B: Confluent hypergeometric function}
Related to the hypergeometric functions $F(a,b,c; z)$, an important role is played in the
theory of special functions by the function
\begin{eqnarray}\label{serie-new}
\Phi(a,b;z)=\sum_{k=0}^\infty\frac{(a)_k}{(b)_k}\frac{z^k}{k!},\,\,\, |z|<\infty,\,\,\, b\neq 0,-1,-2,\cdots
\end{eqnarray}
known as the confluent hypergeometric function. Here $z$ is a complex variable, $a$ and $b$ are parameters
which can take arbitrary real or complex values (except that $b\neq 0,-1,-2,\cdots $), and
\begin{eqnarray}\label{topnew}
(\lambda)_0=1,\,\,\, (\lambda)_k=\frac{\Gamma(\lambda+k)}{\Gamma(\lambda)}=\lambda(\lambda+1)
\cdots (\lambda+k-1),\,\, k=0,1,\cdots.
\end{eqnarray}
The series (\ref{serie-new}) converges for all finite $z,$ and therefore represents an entire function of $z$.
To prove this,  we use the ratio test. Noting that if
\begin{eqnarray}
u_k= \frac{(a)_k}{(b)_k}\frac{z^k}{k!},\mbox{ then }\, \Big|\frac{u_{k+1}}{u_k}\Big|=
\Big|\frac{a+z}{(b+z)(1+k)}z\Big|\rightarrow 0\mbox{ as }
k\rightarrow\infty.
\end{eqnarray}
We can show  that
\begin{eqnarray}
\Phi(a,c;z)=\lim_{b\rightarrow\infty} F(a,b,c;\frac{z}{b}).
\end{eqnarray}
One can also easily check that the confluent hypergeometric function $\Phi(a,b;z)$
is a particular solution of the linear differential
equation
\begin{eqnarray}\label{equKummer}
zf''(z)+(b-z)f'(z)-af(z)=0.
\end{eqnarray}
In fact, denoting the left-hand side of this equation by $L(f)$, and setting $f(z)=f_1(z)=\Phi(a,b;z)$, we have
\begin{eqnarray}
L(f_1(z))&=&\sum_{k=2}^\infty \frac{k(k-1)(a)_k}{(b)_k k!}z^{k-1}+(b-z)\sum_{k=1}^\infty \frac{(a)_k k}{(b)_k k!}z^{k-1}
-a\sum_{k=0}^\infty \frac{(a)_k}{(b)_k k!} z^k\cr
&=&\Big[b\frac{(a)_1}{(b)_1}-a\Big]+\sum_{k=1}^\infty \frac{(a)_k z^k}{(b)_k k!}\Big[k\frac{a+k}{b+k}+b\frac{a+k}{b+k}
-k-a\Big]=0.
\end{eqnarray}
To obtain a second linearly independent solution of (\ref{equKummer}), we assume that $|arg z|<\pi$ and make the substitution
$f(z)=z^{1-b}g(z)$. Then the equation (\ref{equKummer}) goes into an equation of the same form, i.e.,
\begin{eqnarray}\label{equKummernew}
zg''(z)+(b'-z)g'(z)-a'g(z)=0
\end{eqnarray}
with new parameters $a'=1+a-b$, $b'=2-b$. It follows that the function
\begin{eqnarray}
f(z)=f_2(z)=z^{1-z}\Phi(1+a-b, 2-b;z)
\end{eqnarray}
is also a solution of (\ref{equKummer}) if $b\neq 2,3,\cdots.$ Thus, if $b\neq 0,\pm 1,\pm 2,\cdots$, both solutions $f_1$
and $f_2$ are meaningful and are linearly independent, (except for the case $b=1$ where $f_1=f_2$), so that
the general solution of (\ref{equKummer}) can be written in the form
\begin{eqnarray}
f(z)=A\Phi(a,b;z)+B z^{1-b}\Phi(1+a-b,2-b;z),
\end{eqnarray}
where $\,|arg z|<\pi,\,\,b\neq 0,\pm 1,\pm 2,\cdots.$

According to the definition of the Hermite polynomials,
\begin{eqnarray}
H_n(z)=\sum_{k=0}^{[n/2]}\frac{(-1)^kn!}{k!(n-2k)!}(2x)^{n-2k}
\end{eqnarray}
 the even polynomials can be written in the form
\begin{eqnarray}
H_{2n}(z)&=&\sum_{k=0}^n (-1)^k\frac{(2n)!}{k!(2n-2k)!}(2z)^{2n-2k}=(-1)^n(2n)!\sum_{k=0}^n\frac{(-1)^k(2z)^{2k}}{(n-k)!(2k)!}
\cr
&=&(-1)^n\frac{(2n)!}{n!}\sum_{k=0}^n\frac{(-n)_k(2z)^{2k}}{(2k)!}=(-1)^n\frac{(2n)!}{n!}\sum_{k=0}^n
\frac{(-n)_k(z^2)^k}{(\frac{1}{2})_kk!},
\end{eqnarray}
since $(2k)!=2^{2k}(\frac{1}{2})_k k!$, and therefore
\begin{eqnarray}
 H_{2n}(z)=(-1)^n\frac{(2n)!}{n!}\Phi(-n,\frac{1}{2}; z^2).
\end{eqnarray}
For the odd Hermite polynomials, we have the analogous formula
\begin{eqnarray}
 H_{2n+1}(z)=(-1)^n\frac{(2n+1)!}{n!}2z\Phi(-n,\frac{3}{2}; z^2).
\end{eqnarray}
The even Laguerre polynomials can be written in the form
\begin{eqnarray}
L_n^\sigma(z)=\sum_{k=0}^n\frac{\Gamma(n+\sigma+1)}{\Gamma(k+\sigma+1)}\frac{(-z)^k}{k!(n-k)!}=\frac{(\sigma+1)_n}{n!}
\sum_{k=0}^n\frac{(-n)_k z^k}{(\sigma+1)_k k!},
\end{eqnarray}
and hence
\begin{eqnarray}\label{Lag}
 L_n^\sigma(z)=\frac{(\sigma+1)_n}{n!}\Phi(-n,\sigma+1;z).
\end{eqnarray}
Note that relation (\ref{magic}) is obtained by using a novel property of Laguerre polynomials given by
\begin{eqnarray}
n\int_0^\infty\, dz\, e^{-z} z^{\beta}[L_n^\sigma(z)]^2=
 (n+\alpha) \int_0^\infty\, dz\, e^{-z} z^{\beta}[L_{n-1}^\sigma(z)]^2. 
\end{eqnarray}

\end{document}